# Layer controlled orbital selective Mott transition in monolayer nickelate

Byungmin Sohn,[1,2,*,†] Minjae Kim,[3,4*] Sangjae Lee,[1,*] Wenzheng Wei,[1] Juan Jiang,[1] Fengmiao Li,[5] Sergey Gorovikov,[6] Marta Zonno,[6] Tor Pedersen[6], Sergey Zhdanovich[6], Ying Liu,[7] Huikai Cheng,[7] Ke Zou,[5] Yu He,[1] Sohrab Ismail-Beigi,[1] Frederick J. Walker[1], and Charles H. Ahn[1,8]

[1]*Department of Applied Physics, Yale University, New Haven, Connecticut 06520, USA*
[2]*Department of Physics, Sungkyunkwan University, Suwon 16419, Korea*
[3]*Department of Semiconductor Science & Technology, Jeonbuk National University, Jeonju-si 54896, Republic of Korea*
[4]*Korea Institute for Advanced Study, Seoul 02455, Korea*
[5]*Department of Physics and Astronomy, University of British Columbia, Vancouver, British Columbia V6T 1Z1, Canada*
[6]*Canadian Light Source, Saskatoon, Saskatchewan S7N 2V3, Canada*
[7]*Thermo Fisher Scientific, Hillsboro, Oregon 97124, USA*
[8]*Department of Physics, Yale University, New Haven, Connecticut 06520, USA*
*These authors contributed equally to this work.
†Corresponding authors: bsohn@skku.edu

**Abstract**

Dimensionality and electronic correlations are crucial elements of many quantum material properties. An example is the change of the electronic structure accompanied by the loss of quasiparticles when a metal is reduced from three dimensions to a lower dimension, where the Coulomb interaction between carriers becomes poorly screened. Here, using angle-resolved photoemission spectroscopy (ARPES), we report an orbital-selective decoherence of spectral density in the perovskite nickelate $LaNiO_3$ towards the monolayer limit. The spectral weight of the $d_{z^2}$ band vanishes much faster than that of the $d_{x^2-y^2}$ band as the thickness of the $LaNiO_3$ layer is decreased to a single unit cell, indicating a stronger correlation effect for the former upon dimensional confinement. Dynamical mean-field theory (DMFT) calculations show an orbital-selective Mott transition largely due to the localization of $d_{z^2}$ electrons along the c axis in the monolayer limit. This orbital-selective correlation effect underpins many macroscopic properties of nickelates, such as metal-to-insulator transition and superconductivity, where most theories are built upon a $d_{x^2-y^2}$-$d_{z^2}$ two-band model.

**Introduction**

Nickelates are an important family of oxide materials that host a plethora of correlated phenomena, such as metal-to-insulator transitions [1-3], antiferromagnetism [4,5], and more recently, high-temperature superconductivity [6,7]. A common thread through all these elements is the joint contribution to the low-energy electronic structure from both the $d_{x^2-y^2}$ and $d_{z^2}$ states [8]. Compared to the high-$T_C$ cuprates at half-filling, nickelates are generally considered to have a smaller on-site Coulomb interaction U, and the primary energy gap is likely a Mott gap between the lower and upper Hubbard bands, rather than a charge transfer gap between the oxygen ligand band and the upper Hubbard band [8,9]. Thus, understanding how correlation effects manifest within the two bands is of foundational importance to building microscopic models and understanding the properties of nickelates.

Of particular relevance is how the low-energy electronic structure evolves as the system's dimension is reduced. This question has led to efforts to independently engineer the $d_{x^2-y^2}$ and $d_{z^2}$ orbitals [10-15]. Advanced synthesis techniques enable the manipulation of the orbital degrees of freedom in nickelates using atomically precise heterostructures [12,13,16-19] and superlattices [20-22]. Furthermore, due to the recent observation of superconductivity in nickelates [6,7,23-25], which has been suggested to arise from the combination of the $d_{x^2-y^2}$ and $d_{z^2}$ electrons, the orbital-dependent correlation effect in nickelates has become a highly sought-after investigation. Layer-by-layer engineering via epitaxial thin film growth offers an atomically precise route to achieve this goal; however, the challenge in surface preparation has bottlenecked the progress [13,26]. In this study, we use *in-situ* oxygen plasma cleaning on epitaxially grown few-monolayer perovskite nickelates, $LaNiO_3$ (LNO). This approach enables the restoration of high-quality surfaces for ARPES investigation of the intrinsic band properties as the system is tuned to the two-dimensional (2D) limit.

**Results**

The nickel in LNO adopts a $3d^7$ nominal valence shell configuration, where the itinerant electrons near the Fermi level ($E_F$) possess $e_g$ orbital characteristics [3]. Within LNO, the $e_g$ orbital consists of two Ni $3d$ orbitals, $d_{z^2}$ and $d_{x^2-y^2}$. The hopping integrals between the same $d_{z^2}$ orbitals are larger along the out-of-plane z-axis than along the in-plane x- and y-axes (Fig. 1a), whereas the hopping integrals between the same $d_{x^2-y^2}$ are larger along the in-plane directions than along the out-of-plane axis. Hence, as the thickness of LNO decreases to a monolayer, we anticipate distinct behaviors between electrons in $d_{z^2}$ and $d_{x^2-y^2}$ orbitals due to quantum confinement effects [27]. This is likely to induce orbital-dependent modifications to the low-energy band structure as the dimensionality of the LNO is reduced to the 2D limit.

To study the orbital-dependent fermiology of ultrathin LNO, we investigate the electronic structures of $n$ uc (unit-cell) ($n$ = 3, 2, and 1) LNO layers using angle-resolved photoemission spectroscopy (ARPES). Measuring the electronic structures of ultrathin films presents challenges due to extrinsic experimental effects, such as the charging effect, which can often distort the measured spectral weight [28,29]. To address this issue, we initially grow charging-free metallic layers (5 uc LNO layers) on a $LaAlO_3$ (LAO) (001) substrate (Fig. 1b). Subsequently, 8 uc LAO insulating layers, $n$ uc LNO layers, and a 1 uc LAO capping layer are sequentially grown on the 5 uc charging-free LNO layers. It is important to note that the purpose of the 8 uc LAO insulating layers is to electronically isolate the $n$ uc LNO layers, which are of primary interest, from the 5 uc charging-free LNO layers. Additionally, the 1 uc LAO capping layer is grown (Fig. 1c) to protect the $n$ uc LNO layers from potential air contamination and to provide a proper mechanical boundary condition required for the metallicity of the nickelate layer [19,30]. For all films with varying LNO layer thicknesses, the LNO layers are coherently strained to the LAO substrate with ~ 1.4% compressive strain, which rules out the possible effect strain relaxation on the observed electronic structure evolution. (see Fig. S1 in Supplementary Materials (SM)).

Scanning transmission electron microscopy (STEM) images of the single uc nickelate are shown in Fig. 1. In Fig. 1d, a cross-sectional high-angle annular dark-field STEM (HAADF-STEM) image illustrates the $n$ = 1 LNO heterostructure along the LAO [100] zone axis. The LAO (001) substrate, 5 uc LNO charging-free layers, and 8 uc LAO insulating layers are clearly resolved. Energy-dispersive x-ray spectroscopy (EDS) maps for La, Al, and Ni atoms are shown in Fig. 1e-g, respectively. Notably, the 1 uc LNO layer located between the 1 uc capping LAO and 8 uc insulating LAO layers distinctly consists of a nickel monolayer, evident in the Ni EDS intensity map (Fig. 1h). The STEM-EELS measurements for the 1 uc LNO layer shows features of possible intermixing at the interface. However, this is likely due to a well-controlled coverage of the single LNO layer combined with the presence of step-terrace features, which can appear as an intermixing when projected in 2D. Our DFT calculations further suggest that the low-energy electronic structure near the Fermi level from the Ni states is not significantly impacted even in the presence of Al-Ni intermixing (see Fig. S2). With these limitations in mind, we describe below the evolution of the electronic structure of different LNO layer thicknesses arising from intrinsic dimensional confinement effect.

The electronic structures of the $n$ uc LNO layers are characterized using synchrotron-based ARPES. Figures 2a-c show Fermi surface maps of 3, 2, and 1 uc LNO, respectively. A more detailed energy-constant maps are shown in Fig. S3 in SM. In the cases of 3 and 2 uc, two distinct band features are observed at the Fermi surfaces: an electron-like pocket centered at the Γ point, referred to as the α band, and a hole-like pocket centered at the M point, referred to as the β band (Fig. 2d) [31]. However, in 1 uc LNO, the low-energy spectral weights of both bands are suppressed, while a finite portion remains at the Fermi level indicating the metallic nature. This is more clearly illustrated in energy distribution curves (EDCs) depicted in Figs. 2e-g. Note that the EDCs in Figs. 2e-g are extracted from Fig. 2k-m, ensuring that both the EDC and the cuts were measured along the same mirror plane direction. Sharp peaks are observed in 3 and 2 uc LNO, whereas in 1 uc LNO they are significantly suppressed. Interestingly, this decoherence of the spectral density shows a momentum dependence. Figure 2g shows that for the 1 uc LNO, the spectral intensity is more suppressed at the X point than at the zone center (Γ).

To explore the momentum-dependent intensity suppression, we plot energy-momentum (E-$k$) cuts in Figs. 2h-m. First, we focus on the β band. Figures 2h-j are E-$k$ cuts along 'Cut I' represented in Fig. 2a. Momentum-distribution curves (MDCs) along the red dotted line (E = $E_F$) are plotted on top. Two peaks from the β band are well distinguished in MDCs of 3 and 2 uc LNO, whereas a noisy MDC is obtained in 1 uc LNO, which indicates that the β band disappears from $E_F$ in 1 uc LNO. Next, to elucidate how the α band changes with reduced dimensionality, we plot symmetrized E-$k$ cuts along 'Cut II' for each LNO layer thickness (Figs. 2k-m). In 3 and 2 uc LNO, two primary spectral features are observed: i) spectral weight at the Γ point stemming from the α band, and ii) dispersive spectral weight originating from the β band of which the peak positions are $k = \pm 0.6$ Å$^{-1}$ at $E_F$. These two bands are distinctly visible in MDCs at $E_F$, as shown in the upper panels of Fig. 2k and l, with peaks from β bands indicated by black inverted triangles. However, the peaks corresponding to β bands are absent in the 1 uc LNO (upper panel of Fig. 2m), leaving spectral weight only at the Γ point. Although spectral weight is overall suppressed in 1 uc LNO, the suppression is clearly more pronounced for the β band than for the α band (See Fig. S4 in SM for MDCs at other binding energies). Such band-specific adjustments to the fermiology in monolayer LNO suggest an orbital-dependent phenomenon due to the dimensional confinement.

To investigate the origin of such orbital-selective electronic behavior in single uc-thick LNO, we conduct dynamical mean-field theory (DMFT) calculations on LNO layers with thicknesses between 1 and 3 uc (see Method section below and Fig. S5 in Supplementary Materials) confined within LAO layers. For the 1 uc-thick film, the DMFT calculations assume a doping of 0.04 e- per Ni to be consistent with the experimental Fermi surface volume. This is possibly due to extrinsic effects such as oxygen vacancies, or electrostatic doping between the polar [LaO]$^+$ and [NiO$_2$]$^-$ layers. The DMFT-derived Fermi surfaces are presented in Figs. 3a-c along with dispersive electronic structures for each thickness in Figs. 3d-f.

As shown in the X-Γ high symmetry cuts, an electron-like spectral weight exists at the Γ point (from the α band), and spectral weight from β bands (highlighted with red triangles) is obtained near $E_F$. The calculated spectral weights from β bands for the 3 and 2 uc LNO agree well with our ARPES-measured MDCs depicted in the upper panels of Figs. 2k-m. In contrast, for the 1 uc LNO, there is an absence of spectral weight from β bands, and only an electron-like pocket exists at the Γ point (Fig. 3f), which is also consistent with the experiment.

The existence of the single electron-like pocket at the Γ point, with the absence of the spectral weight from the β band, is the signature of the emerging orbital-selective Mott phase. In Figs. 3g-l, we present the orbital-dependent calculated spectral weight for 3 uc (Fig. 3g and 3j), 2 uc (Fig. 3h and 3k), and 1 uc (Fig. 3i and 3l) LNO. The orbital-dependent density of states (DOS) for each LNO thickness are presented in Fig. 3m-o. Our DMFT results show that in 1 uc LNO, only the $d_{x2-y2}$ orbital characterizes the band structure near $E_F$. In contrast, the $d_{z2}$ orbital character band opens a gap and forms (i) a Mott gap having an energy scale of Hubbard $U$ (~ 6 eV, Fig. S2a), and (ii) a Hund satellite peak at E = $E_F$ - 0.3 eV (Fig. 3o) [32-35].

The orbital-selective Mott phase for the 1uc LNO, with the sole presence of the $d_{x2-y2}$ orbital states at $E_F$ (Fig. 3o) and the localized $d_{z2}$ electron, is induced by Hund's coupling. The Hund's coupling creates an additional energy difference between the high-spin (HS) and low-spin (LS) states, where the spins of the electrons in the $d_{x2-y2}$ and $d_{z2}$ orbitals are parallelly or oppositely aligned, respectively (see Fig. S6 for details). The spin-state transition between the multiplet states, whose energetics depend primarily on the Hund's coupling, creates an excitation spectrum below the Fermi level, giving rise to a $d_{z2}$ dominant satellite peak ~0.3 eV below $E_F$ in Fig. 3o. This peak is referred to as a Hund satellite in the literature [34,36]. The presence of the Hund satellite peak implies the importance of the Hund's coupling for the realization of the orbital selective Mott phase [33,37]. Due to the localization of the $d_{z^2}$ orbital, the $d_{z^2}$ spectral weight at $E_F$ proportional to $Z_{z^2}$ ($Z_m$ : quasiparticle residue of the $m$ orbital) diminishes, and the inter-orbital hopping between $d_{x^2-y^2}$ and $d_{z^2}$ is suppressed following $\sqrt{Z_{x^2-y^2} Z_{z^2}} t_{inter-orbital}$. Consequently, the spectral weight from the $d_{z^2}$ orbital dominant β band is suppressed, and the additional doped electron (0.04 e⁻ per Ni) occupies the Γ-centered electron band with a single $d_{x^2-y^2}$ orbital character. This is consistent with the measurement geometry-dependent ARPES results shown in Table S1 and Fig. S7-8, where we find that the orbital character for the α band at the Γ point just below $E_F$ in the energy range of 0.0 – 0.2 eV is switched to the $d_{x^2-y^2}$ orbital for a 1 uc LNO film due to the orbital selective Mott transition.

This result implies that in the orbital-selective Mott phase of 1 uc LNO, the dynamical electronic self-energy effects are essential for the description of the low-energy Fermi surface, with substantial differences from density functional theory results, similar to the orbital-selective Mott phase in iron chalcogenides [32,36]. The feature arising in the 1 uc LNO is different from that in the 2 uc and 3 uc LNO, where the correlation-driven change of the Fermi surface sizes is due to the static self-energy difference of $d_{x^2-y^2}$ and $d_{z^2}$ orbitals in the DMFT (Fig. S9). The unique fermiology of the 1 uc LNO is derived from the interplay between the dimensional confinement, electronic correlation, and the Hund's coupling, and signifies a route to orbital-specific engineering of the electronic structure in strongly-correlated systems. The electronic phases of each layer of LNO are summarized in Table S2.

**Discussion**

In the 3 and 2 uc LNO cases (Fig. 2d), we observe two pockets on the Fermi surface, an electron-like α band centered at Γ and hole-like β bands centered at M. However, in 1 uc LNO, the spectral weight from the hole-like β bands is notably suppressed, leading to a single electron-like band on the Fermi surface, primarily of $d_{x^2-y^2}$ orbital character (Fig. 3c). Our theoretical and experimental observations are summarized in Fig. 4a-c. For thicker LNO layers (n ≥ 2), such as in the cases of 3 and 2 uc, the bandwidth of $d_{z^2}$ ($d_{x^2-y^2}$) orbital character bands is large since the electrons in $d_{z^2}$ ($d_{x^2-y^2}$) orbitals can move along the $z$ direction (in the $xy$ plane) (Fig. 4a). However, in the 1 uc LNO case, where only one $d_{z^2}$ orbital exists along the $z$ axis, electrons in $d_{z^2}$ orbitals can only hop within the $xy$ plane. This restriction leads to a narrowing of the bandwidth of the $d_{z^2}$ orbital character band (Fig. 4b). As a result, when the Coulomb interaction, U, is large enough, the $d_{z^2}$ orbital opens a Hubbard gap (Fig. 4c), with the formation of the Hund satellite peak as shown in Fig. 3o [35,38]. Figure 4d describes how the Hund satellite peak appears due to the inter-site electron hopping. Due to the Hund coupling between $d_{z^2}$ and $d_{x^2-y^2}$ electrons, the Hund satellite peaks appear.

Noteworthy is that the electronic structure of 1 uc LNO differs from the previous theoretical proposal for the $d_{x^2-y^2}$ one orbital Fermi surface of $d^7$ nickelates in Refs. [10,20]. In the previous reports, a tensile strain is considered for the LNO-based superlattices to induce a lower energy level of the $d_{x^2-y^2}$ orbital compared to that of the $d_{z^2}$ orbital, which gives rise to an empty $d_{z^2}$ orbital state. In our 1 uc LNO, the epitaxial strain from the LAO substrate (a = 3.79 Å) induces a compressive strain on LNO (a = 3.86 Å). This compressive strain, accompanied by the elongation of the out-of-plane lattice constant, gives rise to the higher energy level of the $d_{x^2-y^2}$ orbital compared to that of the $d_{z^2}$ orbital (with the local crystal field energy difference of ~440 meV from the Wannier-like orbital in the $e_g$-only model for the projector in the DMFT; see Table S3) [19,39]. As a result, the $d_{z^2}$ orbital is half-filled, and the doped electron occupies the $d_{x^2-y^2}$ orbital due to Hund's coupling as shown in Fig.4c.

The experimental results on 1 uc LNO further suggest the presence of an additional suppression of the spectral weight for the α band, which is beyond the DMFT prediction. Due to the localized nature of the $d_{z^2}$ band in the orbital-selective Mott phase [40], the spin of the $d_{z^2}$ electron can have a strong low-frequency antiferromagnetic spin fluctuation (see Fig. S11 with Refs. [37,41,42]). The spin fluctuation-driven antiferromagnetic alignment of the inter-site $d_{z^2}$ electron's spin blocks the $d_{x^2-y^2}$ electron's hopping due to the energy cost of $J$ (Hund's coupling). This blocking of the $d_{x^2-y^2}$ electron's hopping from the Hund's coupling suppresses the coherence of the $d_{x^2-y^2}$ band, as shown in Fig. 2c.

Note that our DMFT results show that the decoherence of the spectral weight does not depend on the number of electrons in Ni atoms (Fig. S11). Further resonant inelastic x-ray scattering measurements could help verify such low-frequency spin fluctuations expected in the ultrathin LNO layers. Applying a magnetic field to the 1 uc LNO can also restore the coherence of the $d_{x^2-y^2}$ band and enhance the coherence of the Hund satellite peak by suppressing the low-frequency antiferromagnetic spin fluctuation. This spin fluctuation is the source of the dissipation on the inter-site hopping of $d_{x^2-y^2}$ electrons and the HS to LS transitions, inducing the decoherence of the $d_{x^2-y^2}$ band and the Hund satellite.

In summary, we measure the electronic structures of 3, 2, and 1 uc nickelate heterostructures via ARPES and observe an orbital selective Mott phase (OSMP) in a monolayer nickelate system. As the bandwidth of the $d_{z^2}$ orbital band becomes narrow in monolayer LNO due to dimensional confinement [13,16,28,43], an orbital-selective Mott gap opens in the $d_{z^2}$ orbital. The gap opening results in a single band Fermi surface with a single orbital character $d_{x^2-y^2}$ band at the monolayer limit. This work demonstrates the first discovery of the dimensionality-controlled orbital selective Mott phase in nickelates. The interplay between the Hund's coupling and the electronic correlation in the ultrathin nickelates provides us with a rich phase space, including the OSMP emerging at the two-dimensional limit (Table S2). Our findings show the orbital degrees of freedom can be controlled by tuning thin-film thickness, which can offer an orbital-specific tuning knob to investigate the metal-to-insulator transition and superconductivity in nickelate systems. In particular, we believe that our findings are highly significant and timely, especially considering the recent discovery of superconductivity in Ruddlesden–Popper nickelates, where the interplay between orbital selectivity and electron correlations can be closely linked to the emergence of superconductivity [44-46].

**Contributions**

B.S., Y.H., C.H.A., and F.J.W. conceived the project. B.S., S.L., J.J., F.L., K.Z., and Y.H. conducted the ARPES measurements. W.W. prepared and characterized nickelate thin-film samples. Y.L. and H.C. conducted STEM measurements. S.G., M.Z., T.P., and S.Z. developed and maintained the QMSC beamline and ARPES system. M.K. conducted DMFT calculations. B.S., M.K., S.L., Y.H. and F.J.W. wrote the manuscript with input from all the authors.

**Acknowledgments**

We gratefully acknowledge discussions with Younsik Kim and Junsik Mun. Characterization work was supported by the Air Force Office of Scientific Research (AFOSR) under Grant No. FA9550-25-1-0026 and Grant No. FA9550-21-1-0173. Thin film growth was supported by the U.S. Department of Energy, Office of Science, Office of Basic Energy Sciences, under award number DE-SC0019211. Characterization was enabled by a Brookhaven National Laboratory-Yale partner user agreement PU-313536. SIB acknowledges partial salary support from the National Science Foundation (NSF) via grant EAGER DMR 2132343 and thanks the Yale Center for Research Computing for guidance and use of the research computing infrastructure of the Grace high-performance computing cluster. Y.H. acknowledges support from the National Science Foundation under grant DMR2239171. MK was supported by Korea Institute for Advanced Study (KIAS) individual Grants (No. CG083502). The DFT and the DFT+DMFT calculations are supported by the Center for Advanced Computation at KIAS. Part of the research described in this paper was performed at the QMSC beamline at the Canadian Light Source, a national research facility of the University of Saskatchewan, which is supported by the Canada Foundation for Innovation (CFI), the Natural Sciences and Engineering Research Council (NSERC), the Canadian Institutes of Health Research (CIHR), the Government of Saskatchewan, and the University of Saskatchewan.

**Methods**
**Synthesis of nickelate heterostructures**

$n$ ($n$ = 1, 2, and 3) unit cell (uc) LaNiO$_3$ heterostructures are prepared by oxygen-plasma-assisted molecular beam epitaxy on an as-received LaAlO$_3$ (001) substrate from Crystec GmbH. Prior to the growth, the substrate was first cleaned with oxygen plasma at 500 °C for 15 minutes before being heated up to 580 – 700 °C. Then, 5 uc LaNiO$_3$, 8 uc LaAlO$_3$, $n$ ($n$ = 1, 2, and 3) uc LaNiO$_3$, and 1 uc LaAlO$_3$ are sequentially grown on the substrate. The growth rates of LaAlO$_3$ and LaNiO$_3$ are calibrated by a quartz crystal microbalance to 2 Å/min, respectively. The growth was *in-situ* monitored by reflection high-energy electron diffraction (RHEED).

**Angle-resolved photoemission spectroscopy (ARPES)**

ARPES measurements are performed at the Quantum Materials Spectroscopy Centre (QMSC) beamline at the Canadian Light Source with a Scienta R4000 analyzer. A linear horizontally-polarized beam with photon energy of 200 eV is used to measure constant energy maps and energy-momentum distribution cuts. ARPES data are measured at a temperature of 19 K. Before the ARPES measurements, the surfaces of LNO heterostructures are cleaned by using an oxygen plasma source for 10 mins. Then, the LNO heterostructures are transferred into the ARPES chamber.

**Dynamical mean field theory (DMFT)**

Electronic structure calculations in the DFT+DMFT framework [47,48] from the e$_g$ only projector method were performed using the full potential implementation in the TRIQS library [49,50]. The DFT part of the computations in the local density approximation was performed employing the WIEN2k package [51]. We used 18x18x2 k-point mesh for the Brillouin zone integration. Wannier-like e$_g$ orbitals were constructed from the Kohn-Sham bands in the energy range of [-1.2,3.1] eV from the Fermi level. We used the full rotationally invariant Kanamori interaction (ROI) with U = 6.4 eV and J = 0.7 eV. This method of calculation has been suggested to describe the dimensionality-driven evolution of electronic structures of LaNiO$_3$ heterostructures [10,11]. We used a Hubbard parameter, U = 6.4 eV, because it describes the experimentally observed dimensionality-driven Mott transition and the single orbital electronic structure with the orbital selective Mott gap as discussed in the main text (see Table S2). The quantum impurity problem in the DMFT was solved using the continuous-time hybridization-expansion quantum Monte Carlo impurity solver which was implemented in the TRIQS library [52,53]. We perform the one-shot DFT+DMFT computation for the paramagnetic phase, and the temperature is set as 116 K. We confirmed that the electronic structure of the orbital selective Mott phase of 1uc LNO is robust upon lowering the temperature down to 35 K (see Fig. S12).

For the simulation of n uc LNO films, we consider [LNO]$_n$-[LAO]$_5$ superlattices with the lattice constant of a = 3.79 Å, c = 3.94 Å without rotation and tilting, adapted from experiments [19,39]. We confirmed that the 5 uc LAO buffer layer in the computations fully suppresses the out-of-plane dispersion of the e$_g$ orbital driven low energy bands. The electron doping for 1 uc LNO was included in the DMFT part of this calculation, while the DFT step was done for the stoichiometric 1 uc LNO crystal structure. It is essential to note that our assumption involves a doping of 0.04 electrons per Ni atom in the 1 uc LNO, based on our observation of (i) distinct positions of t$_{2g}$ and O$_{2p}$ non-bonding peaks, and (ii) the Hund satellite peak position (See Fig. S6 and Fig. S10 for details). We hypothesize that the (001) polar surface of the LaAlO$_3$ buffer layers may contribute to a slight electron doping effect in the 1 uc nickelate layer, similar to the case previously reported in a cuprate monolayer [54].

**Scanning transmission electron microscopy (STEM)**

A cross-sectional TEM lamella was prepared using Thermo Scientific Helios 5 FX. The region of interest was protected by depositions to obtain the best results, such as curtaining control, etc. Following the TEM sample preparation workflow, the lamella was generated, transferred to a TEM grid with the EasyLift nanomanipulator and then thinned with a focused ion beam (FIB) at 500 V to minimize FIB damage.

High-angle annular dark-field scanning transmission electron microscopy (HAADF-STEM) images and energy-dispersive X-ray spectroscopy (EDS) maps were acquired using a Thermo Scientific

Spectral Ultra S/TEM. The EDS map was acquired at 200 kV with 21 mrad semi-convergence angle of the electron probe and 100 pA probe current for about 6 minutes. An Ultra-X EDS detector with more than 4 srad collection angle ensures obtaining high quality chemical information at an atomic level before sample degradation occurs.

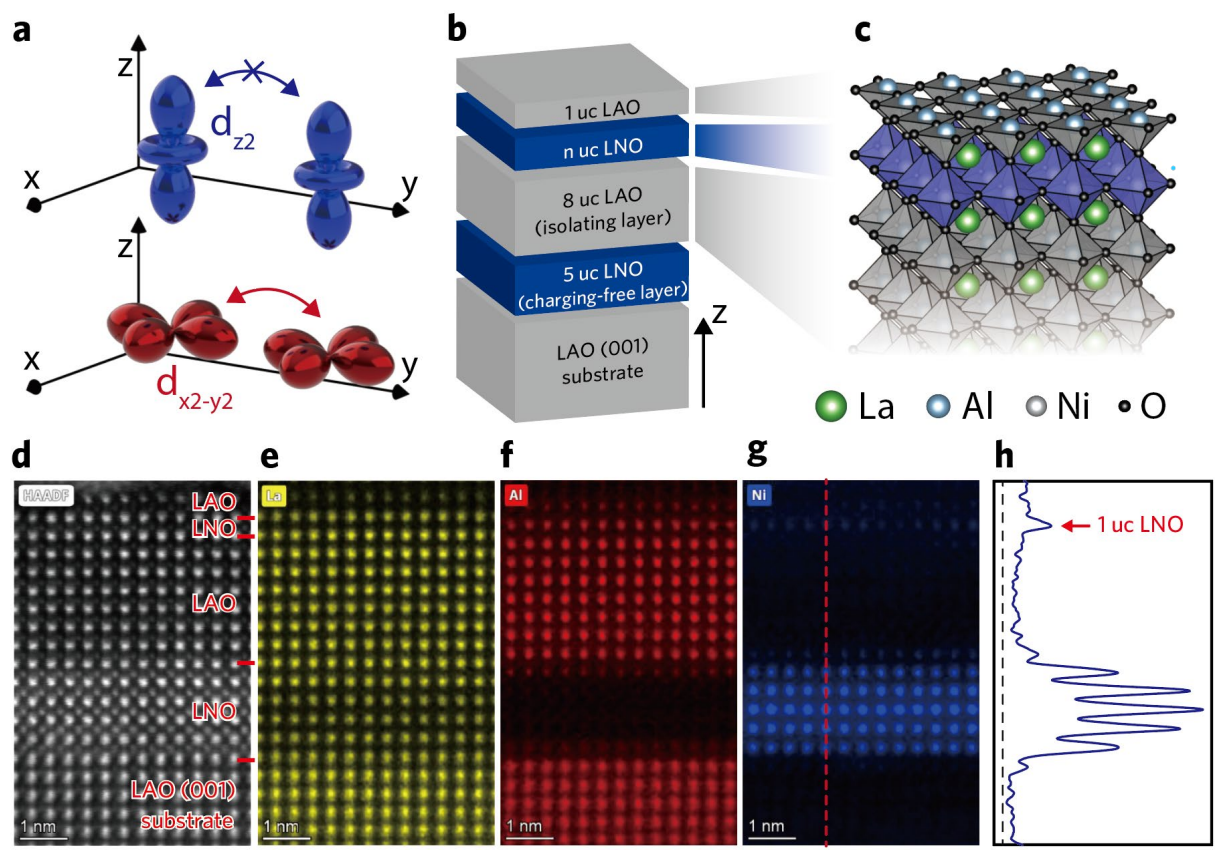

**Figure 1. Dimensionality-controlled atomic-scale nickelate heterostructures.** (a) Schematics of electron hopping integrals between $e_g$ orbitals. (b) A schematic of n (n = 3, 2, and 1) unit-cell (uc) charging-free nickelate heterostructures. LNO (LAO) represents LaNiO$_3$ (LaAlO$_3$). (c) A schematic of the 1 uc charging-free nickelate heterostructure. (d) Atomic-scale imaging of the 1 uc charging-free nickelate heterostructure measured by high-angle annular dark-field scanning transmission electron

microscopy (HAADF-STEM). (e-g) Energy-dispersive x-ray spectroscopy (EDS) maps for La, Al, and Ni. (h) Ni EDS map intensity across a red dotted line in (g).

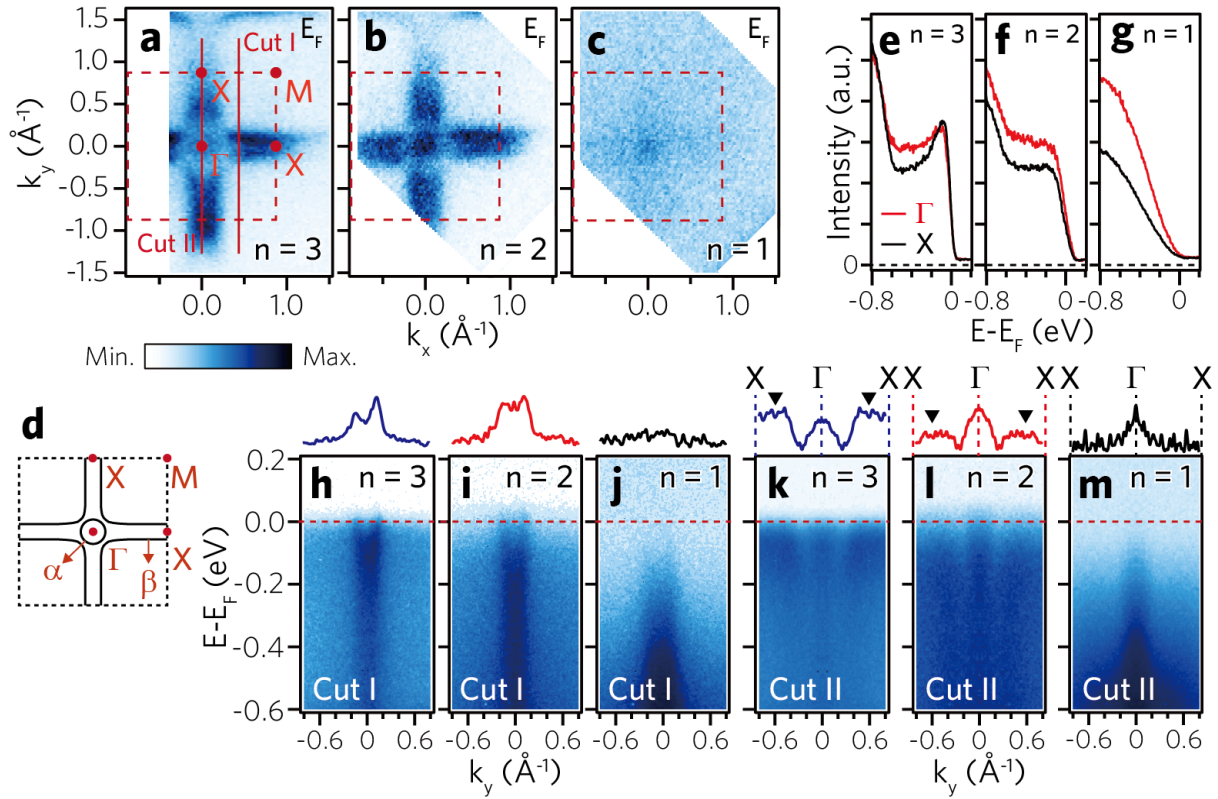

**Figure 2. Electronic structures of the nickelate heterostructures measured by angle-resolved photoemission spectroscopy (ARPES).** (a-c) Fermi surfaces of 3, 2, and 1 uc LNO heterostructures. (d) Fermi surface schematics of 3 and 2 uc LNO heterostructures. (e-g) Energy distribution curves (EDCs) of 3, 2, and 1 uc nickelate heterostructures at the Γ and X points in (k-m), integrated over a range of $\pm 0.075$ Å$^{-1}$. (h-j) Energy-momentum cuts of 3, 2, and 1 uc LNO heterostructures through 'Cut I' in (a). Momentum distribution curves (MDCs) at the Fermi level ($E_F$, red dotted lines) are plotted on top. (k-m) Symmetrized high-symmetry energy-momentum cuts of 3, 2, and 1 uc LNO heterostructures through 'Cut II' in (a). MDCs at $E_F$ (red dotted lines) are plotted on top. Note that the MDCs above (h-m) are integrated over a range of $\pm 30$ meV. Peaks marked with black inverted triangles appear in the 3 and 2 uc heterostructures, whereas the peaks are not observed in the 1 uc heterostructure. All data are obtained at a temperature of 19 K.

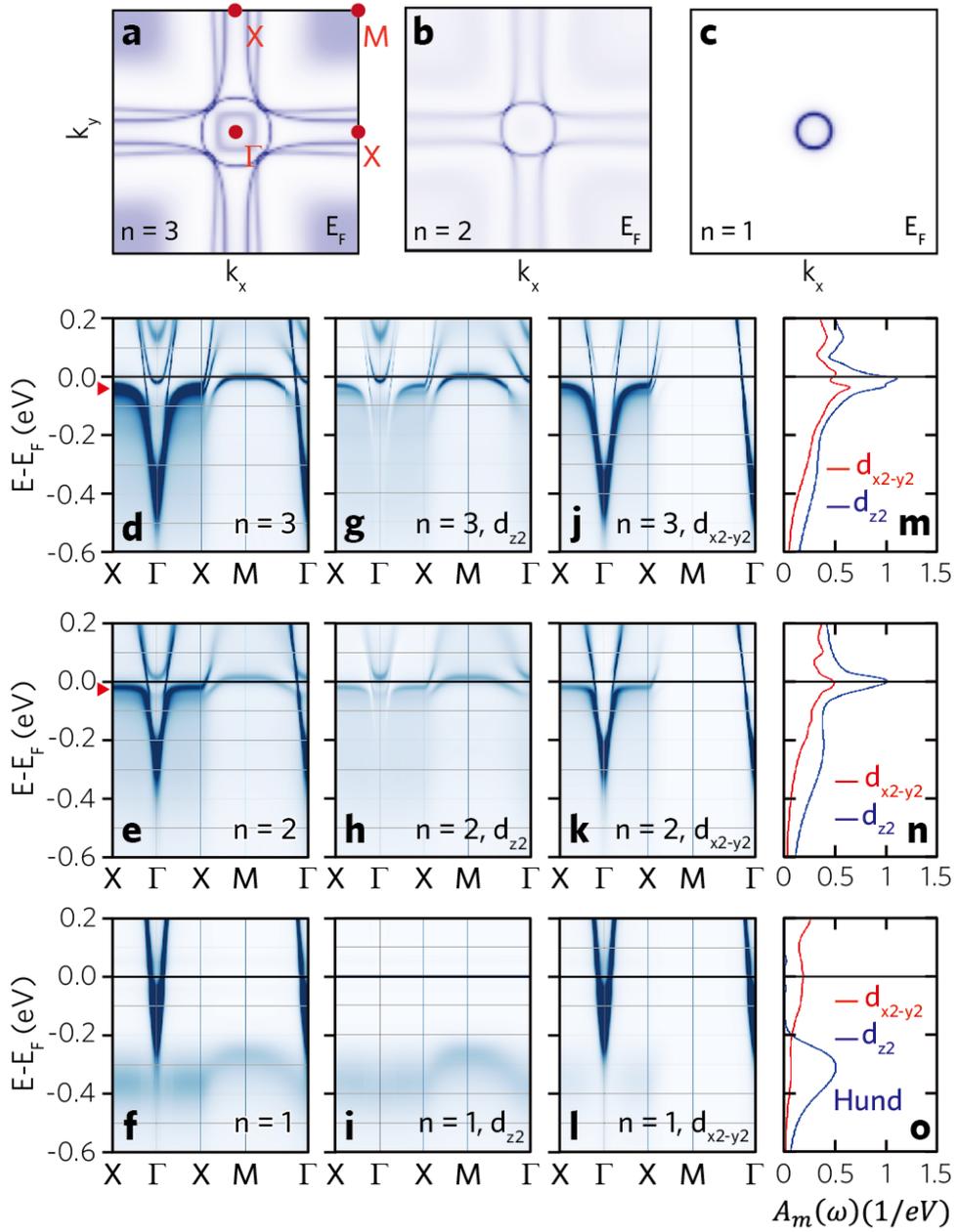

**Figure 3. Calculated ARPES spectra based on dynamical mean-field theory (DMFT).** (a-c) Calculated Fermi surfaces of (a) 3, (b) 2, and (c) 1 uc LNO heterostructures. (d-l) DMFT-calculated E-k spectra of (d) 3, (e) 2, and (f) 1 uc LNO heterostructures. Only $e_g$ orbital-character spectral weights are plotted. (g-l) $d_{z^2}$ and $d_{x^2-y^2}$ orbital-resolved DMFT spectra of a (g,j) 3, (h,k) 2, and (i,l) 1 uc LNO heterostructures are plotted. (m-o) Orbital-dependent calculations of density of states (DOS) of (m) 3, (n) 2, and (o) 1 uc LNO heterostructures are shown. 'Hund' in (o) indicates a Hund satellite peak.

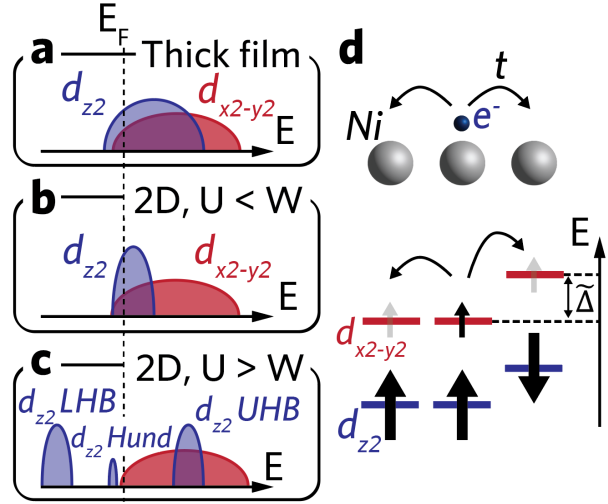

**Figure 4. Single orbital nickel electronic structures in a monolayer LNO.** (a) Schematic illustration of electronic structures of thick LNO thin films. (b,c) Schematic illustrations of electronic structures of a monolayer LNO when the Coulomb interaction energy, U, is (b) small and (c) large. LHB and UHB stand for lower- and upper-Hubbard band, respectively. The Hund satellite appears between LHB and $E_F$. (d) Schematic multiplet configuration of the orbital selective Mott phase in 1 uc LNO. Electrons in Ni $3d_{x^2-y^2}$ orbitals have a hopping constant, $t$. Hund's coupling blocks hopping to the antiferromagnetically correlated neighboring site, causing an additional localization with binding energy of $\tilde{\Delta}$ (Fig. S6 for details).